\documentstyle[aps,eqsecnum,prc]{revtex}

\begin{document}
\draft
\title{ Separation of the monopole contribution to the nuclear Hamiltonian}
\author{ Andr\'es P. Zuker and Marianne Dufour}
\address{ Physique Th\'eorique, B\^at40/1 CRN, IN2P3-CNRS/Universit\'e Louis
Pasteur BP 28, F-67037 Strasbourg Cedex 2, France}
\date{\today}
\maketitle
\begin{abstract}
  It is shown that the nuclear Hamiltonian can be separated rigorously
  as ${\cal H}={\cal H}_m+{\cal H}_M.$ For sufficiently smooth forces
  the monopole part ${\cal H}_m $is entirely responsible for
  Hartree-Fock selfconsistency and hence saturation properties. The
  multipole part ${\cal H}_M$ contains the ``residual'' terms -
  pairing, quadrupole, etc. - that will be analyzed in a companion
  paper. We include a review of the basic results often needed when
  working with multipole decompositions and average monopole
  properties.
\end{abstract}


It has not been possible, yet, to construct interactions that could
satisfy simultaneously three basic conditions:

\begin{itemize}
\item[A)] to be realistic, i.e. consistent with the nucleon-nucleon ($NN$)
 phase shifts,

\item[B)] to ensure good saturation properties, i.e. correct binding energies
at
the observed radii,

\item[C)] to provide good spectroscopy.

\end{itemize}

As a consequence many forces have been designed for specific contexts
or problems : pairing plus quadrupole \cite{BK68,BS69}, density
dependent potentials for mean-field approaches \cite{VB72,DG80},
Landau-Migdal parametrizations for studies of the giant resonances
\cite{GS82}, direct fits to two-body matrix elements for shell-model
calculations \cite{BW88} and many others. A way out of this
unsatisfactory state of affairs would be to exhibit an interaction
consistent with conditions A,B,C above. If we only assume that it
exists and could be reduced to an effective form smooth enough to do
Hartree-Fock (HF) variation, many of its properties can be discovered.
The basic tool we shall need is the following separation property:

{\it Given a sufficiently smooth Hamiltonian ,${\cal H}$, it can be
  separated as ${\cal H}={\cal H}_m+{\cal H}_M.$ Only the monopole
  field ${\cal H}_m$ is affected by spherical Hartree-Fock variation.
  Therefore it is entirely responsible for global saturation
  properties and single particle behaviour.}

As long as we do not find ways of reconciling conditions A and B - a
major problem - $ {\cal H}_m$ must be treated phenomenologically.  The
phenomenological enforcement of good saturation properties is quite
feasible in a shell-model context and leads to the pleasing result
that the multipole part ${\cal H}_M$ - which can be extracted rather
uniquely from the realistic interactions - has an excellent behaviour
\cite{ACZ91,CPZM94}.  Therefore conditions A and C, as well as B and
C, are mutually compatible.

An elementary argument explains the situation.  The observed nuclear
radii, $r\cong 1.2\,A^{1/3}\,$fm, imply average interparticle
distances of some $2.4\,$fm, and therefore the nucleons ``see''
predominantly the medium range of the potential. This is a region that
is well understood theoretically \cite{BJ76} and well described by the
realistic forces.

The picture that emerges is quite simple: the short range part of the
$NN$ interaction is not well understood, but it is certainly
responsible for the repulsive terms necessary for saturation.
Therefore, with our present knowledge of the nuclear forces, the
phenomenological treatment of ${\cal H}_m$ is a necessity, but it may
perhaps lead to some fresh ideas. The argument is that a Fock
representation that can accomodate {\it all} possible non-relativistic
interactions. Although in principle enormously many matrix elements
have to be specified, ${\cal H}_m$ is described by a small fraction of
the total and by some formal manipulations we shall be able to isolate
the {\it very few} that count (essentially the ones describing the
bulk properties of nuclear matter).  Therefore it may possible to
exhibit some simple ${\cal H}_m$ that contains few parameters and
describes nuclear data satisfactorily. Since something simple in Fock
space may be complicated in coordinate space, it may point to ways -
so far overlooked - of reconciling conditions A and B.

There are three steps in the task of giving a complete
characterization of ${\cal H}$.

1. The first is technical: we must prove the separation property, and
prepare the tools that make it possible to take the greatest advantage
of the underlying symmetries: angular momentum,isospin and the unitary
symmetries of the monopole world.

2. As mentioned, we know much about ${\cal H}_M$, but this knowledge
comes in huge arrays of matrix elements. It would be far more useful
if we could extract from this mass of numbers the truly important
ones.

3. It remains to specify the monopole field.

We shall deal with the first point here, and with the second in a
companion (next) paper \cite{DZ} where it is shown that ${\cal H}_M$
is indeed simple and a hint emerges about ${\cal H}_m$. The hint has
been taken up in \cite{DZ95}, a study of nuclear masses that provides
a first approximation to point 3.

\medskip

Our main purpose now is to prove the separation property of the
monopole field . The proof is far more compact if multipole (i.e.
angular momentum) techniques are introduced. Furthermore, they are
essential for the next paper. French lectures \cite{FR66} are the
fundamental reference on the subject and - supplemented by a good book
on angular momentum \cite{BS62} - they contain most of the results we
may need.  In assembling them in the first section and the appendix,
we have tried to give a self-contained account, and compress in a few
pages many of the things of use in daily shell model practice, putting
some emphasis on the connection between coupled and uncoupled
representations, and on questions of phase.

The second section deals with separation and HF properties and it aims
at making the reader familiar with basic technical and conceptual
aspects of the monopole field.

Notations are regrouped at the end of the appendix.

\section{BASIC COUPLINGS AND REPRESENTATIONS}

This section deals with angular momentum coupling and recoupling of
operators following the techniques of French. There is a little here
that is not adapted directly from \cite{FR66}: some changes in
notation, the use of scalar product whenever possible, a lesser
reluctance to go over the $m$-scheme for simple operations and the
introduction of the {\it two} possible analogues of hermitean
conjugation in the coupled schemes.

The reader is reminded that in the Appendix, eqs.(A16-21) contain the
elementary formulae that may be needed in handling $3j$ and $6j$
symbols.

\subsection{ Technical preliminaries}

{\bf The uncoupled representation or $m$-scheme}

We shall work in spaces containing $D$ orbits labeled
$i,\;j,\;k,\;\ell \ldots$ each associated with operators $a_i^+,\;a_i$
that act on a vacuum $| 0>$
\begin{equation}
a_i^+| 0>=| i>\,,\;
a_i| 0>=0\,,\{a_i,a_j^+\}=a_ia_j^++a_j^+a_i=\delta_{ij}
\label{(I.1)}
\end{equation}

For a given number of particles $n$, we can construct $D\choose n$
different states, each associated with a state operator $Z^+$:
\begin{equation}
Z^+([{\bf  i}],n)| 0>=a_{i_1}^+a_{i_2}^+\ldots a_{i_n}^+| 0>=
|[{\bf  i}],n>\,,\;i_1< i_2<\ldots< i_n
\label{(I.2)}
\end{equation}

Operators of rank $k$ take the form (we use * for Hermitean - or
complex - conjugation so as to reserve the + superscript to creation
operators)
\begin{equation}
R_k=\sum_{i,i'}< [{\bf  i}] k| R_k|[{\bf   i'}] k>
Z^+([{\bf  i}],k)Z( [{\bf  i'}],k),\qquad
Z([{\bf  i'}],k)=\left(Z^+([{\bf  i'}] ,k)\right)^*
\label{(I.3)}
\end{equation}

We shall be mainly interested in the $k=1$ case, in the Hamiltonian of
rank $1+2$ (kinetic and potential energies $K$ and $V$ respectively):
\begin{eqnarray}
{\cal H}=K+V =
\sum_{i,j}K_{ij}\,a_i^+a_j-
\sum_{{i< j}\atop {k<\ell}}V_{ijk\ell}\,a_i^+a_j^+a_ka_\ell
\nonumber\\
K_{ij}=< i| K| j>\,;\quad
V_{ijk\ell}=<{ij-ji\over \sqrt2}| V|{k\ell-\ell k\over\sqrt2}>,
\label{(I.4)}
\end{eqnarray}
and in the special $k$-body operator
\begin{equation}
{\cal D}_k=\sum_i
Z^+([{\bf  i}],k)Z([{\bf  i}],k)={n(n-1)\ldots (n-k+1)\over k!}\equiv
{n^{(k)}\over k!}\equiv {n\choose k}
\label{(I.5)}
\end{equation}
whose explicit form is dictated by its rank and the fact that it has
eigenvalue 1 when acting on any state of $k$ particles.

\medskip

{\bf The particle hole (ph) transformations}
\begin{eqnarray}
a_i^+=\bar a_i\,,\qquad a_i=\bar a_i^+\qquad{\rm if}\;i\leq i_f\nonumber\\
a_i^+=\bar a_i^+\,,\qquad a_i=\bar a_i\qquad{\rm if}\;i> i_f
\label{(I.6)}
\end{eqnarray}
consist in interchanging creation and annihilation operators for some
orbits, by interchanging vacua
\begin{equation}
| 0>\;\to\;|\bar 0>=
Z^+([{\bf  i}],f)| 0>\,,\;i\leq i_f\,.
\label{(I.7)}
\end{equation}

In its simplest form, $f=D$ and all operators have to be brought to
normal order. In particular
\begin{eqnarray}
V=-\sum_ {  i< j\atop k<\ell} V_{ijk\ell}
a_i^+a_j^+a_ka_\ell=-\sum V_{ijk\ell}
\bar a_i\bar a_j\bar a_k^+\bar a_\ell^+=
-\sum V_{ijk\ell}[
\bar a_k^+\bar a_\ell^+\bar a_i\bar a_j
\nonumber\\
 +\delta_{i\ell}\delta_{jk}-
\delta_{ik}\delta_{j\ell}-
\bar a_\ell^+\bar a_i\delta_{jk}-
\bar a_k^+\bar a_j\delta_{i\ell}+
\bar a_k^+\bar a_i\delta_{j\ell}+
\bar a_\ell^+\bar a_j\delta_{ik}]
\label{(I.8)}
\end{eqnarray}
and
\begin{equation}
\bar n=D-n\,,\;
\sum Z([{\bf  i}],k) Z^+([{\bf  i}],k)=
\sum \bar Z^+([{\bf  i}],k)\bar Z([{\bf  i}],k)={\bar n\choose k}
\label{(I.9)}
\end{equation}

{\bf Coupled representations}

To take advantage of the basic symmetries, operators have to be
coupled to good angular momentum $J$ and isospin $T$. The $m$-scheme
labels $i,\;j\ldots$ will be replaced by pairs ($r\,r_z)$,
$(s,\,s_z)\,\ldots$ where $r$ specifies the subshell to which the
orbit belongs, and $r_z$ its projection quantum numbers.  Following
French \cite{FR66}, we shall introduce a product notation in which
expressions are independent of the coupling scheme. In $jt$ formalism,
for example, a single tensorial index will represent pairs in
spin-isospin $(JT)$ space
\begin{equation}
\Gamma\equiv JT\,,\quad \Gamma_z\equiv MT_z\,,\quad
r\equiv j_r {1\over 2}\,,\quad  r_z\equiv m_r\,\tau_{z_r}\,,\quad{\rm etc.}
\label{(I.10)}
\end{equation}

(Note that $r$ as tensor index has not exactly the same meaning as
label of shell $r$ (i.e. $r\equiv j_r\,p_r$, where $p_r$ is the
principal quantum number). No confusion can possibly arise from this
convention.)

Expressions involving these indices will stand for products as in
\begin{equation}
(-)^{\Gamma_z}=(-)^{M+T_z},\;
(-)^r=(-)^{j_r+1/2},\;[r]=2(2j_r+1),\;[\Gamma]=[ JT]=
(2J+1)\,(2T+1),
\label{(I.11)}
 \end{equation}

 more generally:
\begin{equation}U(\Gamma{\rm~space})=U(J{\rm~space})U(T{\rm~space})\,,
\label{(I.12)}   \end{equation}

where $U$ may be some $6j$ or Clebsh-Gordan coefficient or similar
functions as in
\begin{equation}< \gamma\gamma_z\,\gamma'\gamma'_z|\Gamma\Gamma_z>=
< jm\,j'm'| JM>\,< tt_z\,t't'_z | TT_z>
\label{(I.13)}
\end{equation}

In $j$ formalism, also called neutron-proton (np), we do not couple
explicitly to good $T$, the tensorial indeces refer to a single space
, and the identifications are $\Gamma=J$, $r\equiv j_r$, $[r]
=(2j_r+1)$ etc. (Note that when used as label, $r$ must specify
whether the shell is a neutron or a proton one).  In $\ell s$ and
$\ell st$ formalisms, we have $(L)(S)$ and $(L)(S)(T)$ product spaces
respectively, and the necessary identifications are straightforward as
long as we do not recouple $L$ and $S$ explicitly to good $J$.

Let us introduce the coupled and {\it normalized} two body state
operators written in terms of {\it ordered} pairs ($a_i^+a_j^+,\,i<
j)$:
\begin{eqnarray}
Z^+_{\Gamma\Gamma_z}(rs)=\zeta_{rs}^{-1}\sum_{(r_z)}
< rr_zss_z|\Gamma\Gamma_z> a_{rr_z}^+ a_{ss_z}^+\,,\quad
Z_{\Gamma\Gamma_z}(rs)=
\left(Z_{\Gamma\Gamma_z}^+(rs)\right)^*
\nonumber\\
a_{rr_z}^+ a_{ss_z}^+=\zeta_{rs}^{-1}\sum_{(\Gamma)}
< rr_zss_z|\Gamma\Gamma_z>
\,Z_{\Gamma\Gamma_z}^+(rs)\,,\quad\zeta_{rs}=(1+\delta_{rs})^{-1/2}
\label{(I.14)}
\end{eqnarray}

$(r_z)$ means that the sum is restricted to $r_z< 0$ if
$\delta_{rs}=1$

$(\Gamma)$ means that $\Gamma$ must be such that $(-)^{\Gamma+2r}=-1$
if $\delta_{rs}=1$

{\it We can always relax the ordering through}

$\zeta_{rs}^{-1}\sum_{(r_z)}=\zeta_{rs}\sum_{r_z}$.

Let us make the following identifications
\begin{equation}
i\equiv (r,r_z),\;j\equiv (s,s_z),\;
k\equiv (t,t_z),\;\ell\equiv (u,u_z),
\label{(I.15)}
\end{equation}

and take advantage of the ordering in eq.(\protect\ref{(I.14)}) to
write directly
\begin{mathletters}
\begin{equation}
V=\sum_{  i< j\atop k<\ell}V_{ijk\ell}\,Z^+(ij)Z(k\ell)=
\sum_{  r\leq s\atop t\leq u,(\Gamma)}
V_{rstu}^\Gamma\,Z^+_\Gamma(rs)\cdot  Z_\Gamma(tu)
\label{(I.16a)}
\end{equation}
\begin{equation}
V_{ijk\ell}=\sum_{(\Gamma)} V_{rstu}^\Gamma\,
\zeta_{rs}^{-1}\zeta_{tu}^{-1}< rr_z ss_z|
\Gamma\Gamma_z>\,< tt_z uu_z|
\Gamma\Gamma_z>
\label{(I.16b)}
\end{equation}
\begin{equation}
V_{rstu}^{\Gamma\Gamma_z}= V_{rstu}^\Gamma=
\sum_{(r_z)(t_z)} V_{ijk\ell}
\zeta_{rs}^{-1}\zeta_{tu}^{-1}< rr_z ss_z|
\Gamma\Gamma_z>\,< tt_z uu_z|
\Gamma\Gamma_z>
\label{(I.16c)}
\end{equation}
\begin{equation}
V_{turs}^\Gamma= V_{rstu}^\Gamma=-(-)^{r+s-\Gamma}
V_{srut}=(-)^{r+s+t+u}
V_{srut}^\Gamma= -(-)^{t+u-\Gamma}V_{rsut}^\Gamma
\label{(I.16d)}
\end{equation}
\end{mathletters}
Rotational and isospin invariance are made manifest through the scalar
products, which in angular momentum manipulations must be rewritten as
zero coupled pairs. There are {\it two} natural ways of doing so,
depending on the order of the couplings:
\begin{mathletters}
\begin{equation}
Z_\Gamma^+\cdot  Z_\Gamma=
 \sum_{\Gamma_z} Z_{\Gamma\Gamma_z}^+  Z_{\Gamma\Gamma_z}
=\sum_{\Gamma_z}Z_{\Gamma\Gamma_z}^+  Z_{\Gamma-\Gamma_z}^\sim
(-)^{\Gamma-\Gamma_z}=[\Gamma]^{1/2}( Z^+Z^\sim)^0, or
\label{(I.17a)}
 \end{equation}
\begin{equation}
Z_\Gamma\cdot  Z_\Gamma^+=
 \sum_{\Gamma_z} Z_{\Gamma\Gamma_z} Z_{\Gamma\Gamma_z}^+
=\sum_{\Gamma_z}Z_{\Gamma\Gamma_z}^-  Z_{\Gamma-\Gamma_z}^+
(-)^{\Gamma+\Gamma_z}=[\Gamma]^{1/2}( Z^-Z^+)^0,
\label{(I.17b)}
 \end{equation}
\end{mathletters}

where $Z$ has been replaced by the most convenient tensor, which is
either $Z^\sim=\widetilde{Z^+}$, the {\it conjugate} of $Z^+$, or
$Z^-=\overline{Z^+}$, the {\it adjoint} of $Z^+$.  The existence of
these operators follows from the commutation rules of $J$ and $T$ with
any irreducible tensor $P_{\gamma_z}^\gamma$:
\begin{equation}
(P_{\gamma_z}^\gamma)^*=(-)^{\gamma+\gamma_z}
\overline {P^{\gamma}_{-\gamma_z}}=(-)^{\gamma-\gamma_z}
\widetilde {P^{\gamma}_{-\gamma_z}}.
\label{(I.18)}
 \end{equation}
 To stress the unavoidable ambiguity in sign, we have taken the
 unusual step of introducing two operators, the adjoint $\bar P$ and
 conjugate $\widetilde P$, that are the inverse of one another:
\begin{equation}
\widetilde {P^\gamma}=(-)^{2\gamma}\overline {P^\gamma},\quad
\widetilde{\widetilde {P^\gamma}}=
\overline{\overline {P^\gamma}}=(-)^{2\gamma} P^\gamma,\quad
\widetilde{\overline {P^\gamma}}= P^\gamma,
\label{(I.19)}
 \end{equation}

 If $(-)^{2\gamma}=1$ the two operators are identical. If
 $(-)^{2\gamma}=-1$, it becomes impossible to ensure - with a single
 definition - the same sign for scalar product and zero coupling of
 operators. Which version we choose depends on the problem at hand.
 For normal ordering as in eq.(\protect\ref{(I.17a)}), and in
 particular
\begin{equation}
n_r=a_r^+\cdot  a_r=[r]^{1/2}(a_r^+a_r^\sim)^0\,,
\label{(I.20)}
\end{equation}

the natural choice of basic tensors is $a_r^+$ and $a_r^\sim$, for
which we adopt the notation of French:
\begin{equation}
A_{rr_z}=a_{rr_z}^+\qquad B_{rr_z}=a_{rr_z}^\sim=(-)^{r+r_z}a_{r-r_z}\,.
\label{(I.21)}
\end{equation}

With antinormal ordering as in eq.(\protect\ref{(I.17b)}), needed in
ph transforms we find

\begin{equation}
Z_\Gamma\cdot  Z_\Gamma^+=[\Gamma]^{1/2}
(Z_\Gamma^-Z_\Gamma^+)^0\equiv[\Gamma]^{1/2}
(\bar Z_\Gamma^+\bar Z_\Gamma^\sim)^0
\label{(I.22)}
\end{equation}
which simply amounts to a notation

\[Z_\Gamma^-=\bar Z_\Gamma^+,\;Z_\Gamma^+=\bar Z_\Gamma^\sim,\]

as the bar over the operators means ph transform.  But it also means
adjoint, and indeed:

$\overline {Z_\Gamma^+}= Z_\Gamma^-$ by {\it definition} and
$\overline {Z^\sim}= \overline {\widetilde Z^+}=Z^+$ by definition of
$Z^\sim$ and by the last equality in (\protect\ref{(I.19)}). Therefore
the ph transform of a state operator is its adjoint. The antitransform
of a state operator is its conjugate by exactly the same argument.
Note that for the uncoupled operators ph transform is Hermitean
conjugation, better represented by its dual analogues in
(\protect\ref{(I.18)}) than by the arbitrary choice of one of them.
{}From (\protect\ref{(I.22)}) we have the basic transforms:
\begin{equation}
\bar A_{rr_z}=a_{rr_z}^-=(-)^{2r}B_{rr_z}
\qquad \bar B_{rr_z}=A_{rr_z}\,.
\label{(I.23)}
\end{equation}

{\it The coupled operators} quadratic in $A$ and $B$ are
\begin{equation}
X_{\Gamma\Gamma_z}^+(rs)=(A_rA_s)^\Gamma_{\Gamma_z},\quad
X_{\Gamma\Gamma_z}(rs)=(B_rB_s)^\Gamma_{\Gamma_z},\quad
S^\gamma_{\gamma_z}(rt)=(A_rB_t)^\gamma_{\gamma_z}
\label{(I.24)}
\end{equation}

Obviously $Z_{\Gamma\Gamma_z}^+(rs)=\zeta_{rs}
X_{\Gamma\Gamma_z}^+(rs)$. From the easily proved identity
\begin{equation}
(\overline {P^\gamma Q^{\gamma'}})^\Gamma=(-)^{\gamma+\gamma'-\Gamma}
(\bar Q^{\gamma'} \bar P^\gamma)^\Gamma
\label{(I.25)}
\end{equation}

(equally valid for conjugation), we obtain
\begin{equation}
\overline {X_\Gamma^+}(rs)=-X_\Gamma(rs)\qquad
\overline {S_{\gamma_z}^\gamma}(rt)
=(-)^{t-r-\gamma}S_{-\gamma_z}^\gamma(tr)\,.
\label{oubli}
\end{equation}

For reduced matrix elements we use Racah's definition
\begin{equation}
<\alpha\alpha_z| P_{\gamma_z}^\gamma| \beta\beta_z>=
(-)^{\alpha-\alpha_z}\pmatrix{\alpha&\gamma&\beta\cr
-\alpha_z&\gamma_z&\beta_z\cr}\,< \alpha\|  P^\gamma\| \beta>\,.
\label{(I.26)}
\end{equation}

For any operator $P^\gamma$ it is true that
\begin{equation}
<\alpha\alpha_z| P_{\gamma_z}^\gamma| \beta\beta_z>=
<\beta\beta_z| (P_{\gamma_z}^\gamma)^*| \alpha\alpha_z>^*
\label{(I.27)}
\end{equation}

and by applying (\protect\ref{(I.26)}) to both sides it follows that
\begin{equation}
<\alpha \|  P^\gamma\|  \beta>=(-)^{\alpha-\beta-\gamma}
<\beta \|  \bar P^\gamma\|  \alpha>\,,
\label{(I.28)}
\end{equation}

where we have omitted complex conjugation on the rhs because our
reduced matrix elements will be real.

\medskip

The coupled form of a rank 1 operator is deduced from the uncoupled
one in (\protect\ref{(I.3)}) by using (\protect\ref{(I.26)}) and the
definition of $S^\gamma$ in (\protect\ref{(I.24)}):
\begin{equation}
R_{\gamma_z}^\gamma=\sum_{r,t,r_z}< rr_z|
R_{\gamma_z}^\gamma| tt_z> a_{rr_z}^+a_{tt_z}=\sum_{rt}
< r\|  R^\gamma \|  t>(
\gamma)^{-1/2}S_{\gamma_z}^\gamma(rt)\,.
\label{(I.29)}
\end{equation}

We can always rewrite an arbitrary $R^\gamma$ in terms of the
symmetric $({\cal S})$ and antisymmetric $({\cal A})$ operators
\begin{equation}
{\cal S}_{\gamma_z}^\gamma(rt)=S_{\gamma_z}^\gamma(rt)+(-)^{r-t}
S_{\gamma_z}^\gamma(tr),\quad
{\cal A}_{\gamma_z}^\gamma(rt)=S_{\gamma_z}^\gamma(rt)-(-)^{r-t}
S_{\gamma_z}^\gamma(tr),
\label{(I.30)}
\end{equation}

Then calling $R^\gamma_{rt}= < r\| R^\gamma \| t>( \gamma)^{-1/2}$,
the expansion becomes
\begin{equation}
R_{\gamma_z}^\gamma={1\over 2}\sum_{r\leq t}\left\lbrack
(R^\gamma_{rt}+(-)^{r-t}R^\gamma_{tr})
{\cal S}_{\gamma_z}^\gamma(rt)+(R^\gamma_{rt}-(-)^{r-t} R^\gamma_{tr})
{\cal A}_{\gamma_z}^\gamma(rt)\right\rbrack
\label{(I.31)}
\end{equation}

The $(-)^{r-t}$ phase ensures that the spherical harmonics are
symmetric if $Y^*_{\ell m}=(-)^m Y_{\ell-m}$ (Condon and Shortley's
choice) , which leads to $<r\|Y_\ell\|t>\;=\;(-)^{r-t}<t\| Y_\ell\|
r>$ and vanishing of the ${\cal A}$ term. The phase convention should
be changed to $(-)^{r-t-\ell}$ if $Y_\ell\;\to\;i^\ell Y_\ell$. Then
$Y^*_{\ell m}=(-)^{\ell+m} Y_{\ell-m}$, ~and~ $<r\| Y_\ell\| t>=
(-)^{r-t-\ell}<t\| Y_\ell\| r>$.  Note that now $Y_\ell\cdot Y_\ell=
[\ell]^{1/2}(Y_\ell Y_\ell)^0$ instead of the usual $Y_\ell\cdot
Y_\ell=(-)^\ell [\ell]^{1/2}(Y_\ell Y_\ell)^0$ but $J\cdot
J=-\sqrt3(JJ)^0$ and $T\cdot T=-\sqrt3(TT)^0$ always, since $J\to iJ$
makes no sense. Still, positive definite zero coupling for tensors of
integer rank could be obtained with the change $<\ell m\ell-m|
00>=(-)^{\ell-m} [\ell]^{-1/2}$ to $(-)^m[\ell]^{-1/2}$. As we have
seen, for half integer rank, no convention will ensure that zero
coupling is always definite positive.

\medskip
{\bf Recoupling}.

Let us consider a ph transformation (\protect\ref{(I.6)}) and using
the identification (\protect\ref{(I.15)}) assume that orbits $j\equiv
(ss_z)$ and $k\equiv (tt_z)$ are transformed while $i$ and $\ell$
(i.e. $r$ and $u$) are left untouched. Then we have
\begin{equation}
V=\sum_{i< j\atop k<\ell}V_{ijk\ell}\,a_i^+a_ka_j^+a_\ell-
\sum V_{ijj\ell}a_i^+a_\ell\delta_{jk}.
\label{(I.32)}
\end{equation}
Now use (\protect\ref{(I.16b)}) , relax the ordering for projections
in the contraction in (\protect\ref{(I.32)}) , and introduce the $A~B$
tensors, permuting the middle ones

\begin{eqnarray}
V=\sum  V_{rstu}^\Gamma  [\Gamma]^{1/2}
\zeta_{rs}\zeta_{tu}\biggl[
\left\{\left(A_r\Bigl[B_tA_s\Bigr)^{\Gamma}B_u\right]^{\Gamma}\right\}^0
-\nonumber\\
\sum_{\Gamma_{z}s_z}
< rr_z ss_z|\Gamma\Gamma_z>\,< ss_z uu_z|\Gamma\Gamma_z>
 a_{rr_z}^+a_{uu_z}\Biggr].
\label{(I.33)}
\end{eqnarray}

For clarity the internal couplings are indicated by different
notations: $(\;)^\Gamma$ and $[\;]^\Gamma$. Recoupling through a
normalized $9j$ symbol yields

\begin{equation}
\left\{\left(A_r\Bigl[B_tA_s\Bigr)^{\Gamma}B_u\right]^{\Gamma}\right\}^0
=\sum_{\gamma}[\Gamma\gamma]
\left\{\begin{array}{ccc}
r&s&\Gamma\\t&u&\Gamma\\\gamma&\gamma&0
\end{array}\right\}
(S_{rt}^\gamma S_{su}^\gamma)^0.
\label{(I.34)}
\end{equation}
The contraction in (\protect\ref{(I.33)}) can be calculated to be
\begin{equation}
\sum_{\Gamma_{z} s_z}
< rr_z ss_z|
\Gamma\Gamma_z>\,< uu_z ss_z|
\Gamma\Gamma_z>  (-)^{u+s-\Gamma}=\hat\delta_{ur}\delta_{u_zr_z}
(-)^{u+s-\Gamma}\left\lbrack {\Gamma\over r}\right\rbrack^{1/2}\,,
\label{(I.35)}
\end{equation}

obtained by reorganizing the couplings with the help of 3-$j$ symbols.
All these operations can be summed up by

\begin{eqnarray}
-(X_\Gamma^+(rs)X_\Gamma(tu))^0=
\sum_\gamma [\Gamma\gamma)]^{1/2}
(-)^{s+t-\gamma-\Gamma}
\left\lbrace\matrix{r\;s\;\Gamma\cr u\;t\;\gamma\cr}\right\rbrace
(S_{rt}^\gamma S_{su}^\gamma)^0\nonumber\\
-(-)^{u+t-\Gamma}\left\lbrack {\Gamma\over r}\right\rbrack^{1/2}
\delta_{st} S_{ru}^0
\label{(I.36)}
\end{eqnarray}

and its inverse
\begin{eqnarray}
(S_{rt}^\gamma S_{su}^\gamma)^0=
-\sum_\Gamma [\Gamma\gamma)]^{1/2}
(-)^{s+t-\gamma-\Gamma}
\left\lbrace\matrix{r\;s\;\Gamma\cr u\;t\;\gamma\cr}\right\rbrace
(X_\Gamma^+(rs)X_\Gamma(tu))^0\nonumber\\
+(-)^{u-t+\gamma}\left\lbrack {\gamma\over r}\right\rbrack^{1/2}
\delta_{st} S_{ru}^0
\label{(I.37)}
\end{eqnarray}

where the first term can be written by inspection but the contraction
needs some care.  The hybrid procedure we have adopted - of mixing
$a^+,\,a$ and $A,\,B$ - points to a paradox: French's notation and
techniques simplify many complex coupling problems but complicate a
few simple ones. (Hint: try to obtain (\protect\ref{(I.36)}) and
(\protect\ref{(I.37)}) using the - impressive - artillery in Hsu's
appendix to \cite{FR66}; a good exercise).

\subsection {The V and $\omega$ representation
of ${\cal H}$}
\label{I.2}

Equations (\protect\ref{(I.36)}) and (\protect\ref{(I.37)}) were
derived to prepare for a ph transform of some orbits. However, beyond
the possibility of changing vacua - which may be quite useful
occasionally - we are interested in the different representation(s) of
the Hamiltonian that these operations entail.  Accordingly, we shall
always keep the untransformed notation $a_r^+,\;a_r$ and $A_r\;B_r$
for the orbits.

First we consider the case in which all orbits are transformed.
Introducing $S_{ru}=a_r^+\cdot a_u$, the coupled version of
eq.(\protect\ref{(I.8)}) becomes,

  \begin{eqnarray}
V=-\sum_{  r\leq s\atop   t\leq u,\Gamma} V_{rstu}^\Gamma
Z_\Gamma(tu)\cdot  Z^+_\Gamma(rs)-
\sum_{r\leq s,\Gamma}[\Gamma] V_{rsrs}^\Gamma\label{(I.38)}\\
-\sum_{  r\leq s\atop   t\leq u,\Gamma}\zeta_{rs}\zeta_{tu}
[\Gamma] V_{rstu}^\Gamma  \left\lbrack
(-)^{r+s-\Gamma}\left( {S_{ru}\over [r]}\delta_{st}
+ {S_{st}\over [s]}\delta_{ru}\right)-
{S_{rt}\over [r]}\delta_{su}-
{S_{su}\over [s]}\delta_{rt}\right\rbrack
\nonumber
\end{eqnarray}

All contractions are calculated using (\protect\ref{(I.35)}) .

Next we examine the transformations associated to
(\protect\ref{(I.36)}) and (\protect\ref{(I.37)}) in which only the
middle operators are interchanged. It is convenient to allow for
complete flexibility in the summations and we shall relax the
restrictions $r\leq s,\;t\leq u$ by replacing the $\zeta_{rs}$ factors
by the $P_{r s}$ convention
\begin{equation}
P_{r s}=(1+\delta_{rs})^{-1/2}\quad{\rm if}~r\leq s \quad,
P_{r s}={(1+\delta_{rs})^{1/2}\over 2}~{\rm if~no~restriction}
\label{(I.39)}
\end{equation}

so that the sums could be interpreted as restricted or not restricted.
We write therefore
\begin{equation}
V=\sum_{  r\leq s\atop t\leq u,\Gamma} V_{rstu}^\Gamma
Z_{rs\Gamma}^+\cdot  Z_{tu\Gamma}=-
\sum_{(rstu)\Gamma}P_{rs}P_{tu}
[\Gamma]^{1/2} V_{rstu}^\Gamma
(X_{rs\Gamma}^+  X_{tu\Gamma})^0
\label{(I.40)}
\end{equation}
and from now on we set
$Z_\Gamma^+(rs)= Z_{rs\Gamma}^+$ etc.

According to (\protect\ref{(I.36)}), $V$ can be transformed into
\begin{equation}
V=\sum_{(rstu)\gamma}P_{rs}P_{tu}\left\lbrack
[\gamma]^{1/2} \omega_{rtsu}^\gamma
(S_{rt}^\gamma S_{su}^\gamma)^0+\delta_{st}\hat\delta_{ru}
[s]^{1/2} \omega_{russ}^0 S_{ru}^0\right\rbrack
\label{(I.41)}
\end{equation}

where
\begin{mathletters}
\begin{equation}
\omega_{rtsu}^\gamma=
\sum_{(\Gamma)}
(-)^{s+t-\gamma-\Gamma}
\left\lbrace\matrix{r\;s\;\Gamma\cr u\;t\;\gamma\cr}\right\rbrace
V_{rstu}^\Gamma[\Gamma]
\label{(I.42a)}
\end{equation}
\begin{equation}
V_{rstu}^\Gamma=
\sum_{\gamma}
(-)^{s+t-\gamma-\Gamma}
\left\lbrace\matrix{r\;s\;\Gamma\cr u\;t\;\gamma\cr}\right\rbrace
\omega_{rtsu}^\gamma[\gamma]
\label{(I.42b)}
\end{equation}
\end{mathletters}
(Remember that $\sum_{(\Gamma)}$ means that we sum over Pauli allowed
$\Gamma$).  \medskip Eq.(\protect\ref{(I.37)}) suggests an alternative
to (\protect\ref{(I.41)})
\begin{equation}
V=\sum_{(rstu)\gamma}P_{rs}P_{tu}
[\gamma]^{1/2} \omega_{rstu}^\gamma
\left\lbrack
(S_{rt}^\gamma S_{su}^\gamma)^0-(-)^{\gamma+r-s}\left\lbrack
{\gamma\over r}\right\rbrack^{1/2}\delta_{st}\hat\delta_{ru}S_{ru}^0
\right\rbrack
\label{(I.43)}
\end{equation}

where each multipolarity $\gamma$ is associated with a two body
operator.  The obvious check that (\protect\ref{(I.43)}) is indeed
(\protect\ref{(I.41)}) comes from
\begin{equation}
-\sum_\gamma
[\gamma]^{1/2} \omega_{rstu}^\gamma
(-)^{\gamma+r-s}
\left\lbrack{\gamma\over r}\right\rbrack^{1/2}=[s]^{1/2}
\omega_{russ}^0\,.
\label{(I.44)}
\end{equation}

The proof is left as an exercise (use (\protect\ref{(I.42a)}) and
Racah sum rule (A.19)).  \medskip

In a $jt$ representation, by introducing explicitly the isospin in
$\Gamma=JT,\;\gamma=\lambda\tau,\;r=j_r {1\over 2}$ etc and the 6-$j$
values
\begin{equation}
\left\lbrace\matrix{1/2\;1/2\;T\cr
1/2\;1/2\;\tau\cr}\right\rbrace =\matrix{T\tau&00&01&10&11\cr
&-1/2&1/2&1/2&1/6\cr &\phantom{1}&&&\cr}
\label{(I.45)}
\end{equation}

we find
\begin{mathletters}
\begin{equation}
\omega_{rtsu}^{\lambda 0}={1\over 2}
\sum_{(J)}
(-)^{j_s+j_t-\lambda-J}
\left\lbrace\matrix{j_r\;j_s\;J\cr j_u\;j_t\;\lambda\cr}\right\rbrace
[J] (V_{rstu}^{J 0}+3V_{rstu}^{J 1})
\label{(I.46a)}
\end{equation}
\begin{equation}
\omega_{rtsu}^{\lambda 1}={1\over 2}
\sum_{(J)}
(-)^{j_s+j_t-\lambda-J}
\left\lbrace\matrix{j_r\;j_s\;J\cr j_u\;j_t\;\lambda\cr}\right\rbrace
[J] (V_{rstu}^{J 0}-V_{rstu}^{J 1})
\label{(I.46b)}
\end{equation}
\end{mathletters}
and reciprocally
\begin{mathletters}
\begin{equation}
V_{rstu}^{J 0}={1\over 2}
\sum_{\lambda}
(-)^{j_s+j_t-\lambda-J}
\left\lbrace\matrix{j_r\;j_s\;J\cr j_u\;j_t\;\lambda\cr}\right\rbrace
[\lambda] (\omega_{rtsu}^{\lambda 0}+3\omega_{rtsu}^{\lambda 1})
\label{(I.47a)}
\end{equation}
\begin{equation}
V_{rstu}^{J 1}={1\over 2}
\sum_{\lambda}
(-)^{j_s+j_t-\lambda-J}
\left\lbrace\matrix{j_r\;j_s\;J\cr j_u\;j_t\;\lambda\cr}\right\rbrace
[\lambda] (\omega_{rtsu}^{\lambda 0}-\omega_{rtsu}^{\lambda 1})
\label{(I.47b)}
\end{equation}
\end{mathletters}

When the Hamiltonian is written as in (\protect\ref{(I.40)}) we speak
of the normal or $V$-representation while we call the form
(\ref{(I.41)}) multipole, or $\omega$-representation.

\section{ THE MONOPOLE FIELD $  {\cal H}_m$ AND THE SEPARATION
  ${\cal H}={\cal H}_m+{\cal H}_M$}
\label{II}
Before going into the technical problems, it is worth explaining how
the monopole field appears. Several of the statements that follow need
a formal proof. It will be supplied in the body of the section.

The separation of the Hamiltonian into an ``unperturbed'' and a
``residual'' part , ${\cal H}={\cal H}_0+{\cal H}'$, is at the heart
of many-body physics, and the idea that ${\cal H}_0$, could be
represented by some central - i.e. single particle - field is of great
heuristic and qualitative value. However, to decide whether it makes
sense quantitatively, we must understand how ${\cal H}$, of rank-2,
can be approximated by a rank-1 operator.

What can be done cleanly is to define ${\cal H}_0$ as $K$ plus a two
body part so that
\begin{equation}
{\cal H} =K+\sum_{r\leq s}   V_{rs}\,n_r(n_s-\delta_{rs})
             /(1+\delta_{rs})+ {\cal H}'.
\label{new}
\end{equation}
${\cal H}_0$ has two equivalent properties:

{\em Complete Extraction.} When written in a multipole representation
${\cal H}$' contains no number operators. It means that all the
$\lambda=0$ terms that are diagonal in the basis we are using have
migrated to ${\cal H}_0$.

{\em Trace Equivalence.} The expectation value of ${\cal H}_0$ for any
basic state is the average energy (i.e. the trace) of the
configuration to which it belongs. (A configuration is the set of
states with fixed number of particles in each orbit).

To recover and examine the notion of central field we choose as vacuum
some determinantal state, i.e. we set $n_r=D_r$ below the Fermi orbit
and $n_r=0$ above. The single - particle and single - hole energies
are then calculated from eq.(\ref{new}), by adding or removing a
particle from the vacuum.  In a HF calculation these energies play an
important role and invite the interpretation that the determinantal
state generates a mean field in which the particles move. This
interpretation is very good when there is a dominant agent in the
Hamiltonian that is responsible for most of the field, as in atoms and
planets.

In nuclei it is simply wrong. What the HF calculation produces is a
basis of orbitals. When ${\cal H}$ is written in that basis, ${\cal
  H}_0$ can be extracted, and it will yield indeed the HF value for
the vacuum and single fermion energies. However, there is no reason to
{\it linearize} ${\cal H}_0$ (i.e. to approximate it by a central
field) when estimating energies of other configurations: the quadratic
effects grow fast. To fix ideas: typically $V_{rs}\approx300$ keV in
the pf shell, not a large number, but it multiplies the $n_rn_s$
operator that may become $O(10)$ or even $O(100)$, and drastically
modify the effects of the central field.

There is much empirical evidence that at fixed total number of
particles it is a good approximation to keep the same basis for all
orbits in the vicinity of the Fermi level, and therefore ${\cal H}_0$
is fixed. However we also know that, when we change the number of
particles, the orbitals and therefore ${\cal H}_0$ must evolve.

Since the evolution of the orbitals is associated to unitary
transformations of the basis, the $n_r$ operators become linear
combinations of $S^0_{tu}$ ones. Therefore we must generalize the
definition of ${\cal H}_0$, by extending the notion of ``complete
extraction'' to all $\lambda=0$ operators, not only the diagonal ones.
The resulting object is what we call ${\cal H}_m$.  As the separation
${\cal H}={\cal H}_m+{\cal H}_M$ is invariant
(representation-independent), ${\cal H}_m$ provides the -
mathematically and physically - natural definition of unperturbed
Hamiltonian.

\subsection{ Separation of ${\cal H}_m$}

Although there is only one ${\cal H}_m$, it will have different
aspects in $jt$ and $j$ formalisms. We differenciate them by a tag:
\[{\cal H}_m={\cal H}_{mT}={\cal H}_{mnp},\]
${\cal H}_{mT}$ is constucted by extracting all the possible
$\gamma=00$ and $01$ contributions to eqs.(\ref{(I.41)}) or
(\ref{(I.43)}), while ${\cal H}_{mnp}$ contains all the possible
$\gamma=0$ terms. Obviously the kinetic energy is part of ${\cal
  H}_m$, whose rank~2 component must have the form of the $\lambda=0$
contributions to (\ref{(I.43)}) which we call $V_{\lambda=0}$. To give
them a transparent aspect we replace $S_{rs}^{00}$ and $S_{rs}^{01}$
by
\begin{equation}
S_{rs}=\hat\delta_{rs}[r]^{1/2}S_{rs}^{00}\,,\qquad
T_{rs}={1\over 2}\hat\delta_{rs}[r]^{1/2}S_{rs}^{01}
\label{(II.1)}
\end{equation}

which, for $\delta_{rs}=1$, reduce to $S_{rr}=n_r,\;T_{rr}=T_r$.  The
two body operators

\begin{mathletters}
\begin{equation}
S_{rtsu}=\zeta_{rs}\zeta_{tu}(S_{rt}S_{su}-\delta_{st}S_{ru})
\label{(II.2a)}
\end{equation}
\begin{equation}
T_{rtsu}=\zeta_{rs}\zeta_{tu}(T_{rt}\cdot  T_{su} - {3\over 4}
\delta_{st}S_{ru})
\label{(II.2b)}
\end{equation}
\end{mathletters}

in turn become for $\delta_{rt}=\delta_{su}=1$,
\begin{mathletters}
\begin{equation}
S_{rrss}=n_r(n-\delta_{rs})/(1+\delta_{rs})\,,
\label{(II.a')}
\end{equation}
\begin{equation}
T_{rrss}=(T_r\cdot  T_s-{3\over 4}n_r\delta_{rs})/(1+\delta_{rs})\,.
\label{(II.b')}
\end{equation}
\end{mathletters}

and the $\lambda=0$ contribution to $V$ can be written as
\begin{equation}
V_{\lambda=0}=
\sum_{  r\leq s,\atop t\leq u}[rs]^{-1/2}
(\omega_{rtsu}^{00}S_{rtsu}-4\omega_{rtsu}^{01}T_{rtsu})
\hat\delta_{rt}\hat\delta_{su},
\label{(II.3)}
\end{equation}

which follows directly from (\ref{(II.2a)}), (\ref{(II.2b)}) and
(\ref{(I.43)}) provided we remember that $T_{rt}\cdot T_{su}=-\sqrt
3(T_{rt}T_{su})^0$.

Although the full rank-2 contribution to ${\cal H}_{mT}$ has the form
$V_{\lambda=0}$, the correct values of the $\omega_{rtsu}^{00}$ and
$\omega_{rtsu}^{01}$ cannot be extracted from eqs.(\ref{(I.46a)}) and
(\ref{(I.46b)}) as
\begin{mathletters}
\begin{equation}
\omega_{rtsu}^{00}=[rs]^{-1/2}
\sum_{(J)} [J]
(V_{rstu}^{J 0}+3V_{rstu}^{J 1})
\hat\delta_{rt}\hat\delta_{su}
\label{(II.4a)}
\end{equation}
\begin{equation}
\omega_{rtsu}^{01}=[rs]^{-1/2}
\sum_{(J)} [J]
(V_{rstu}^{J 0}-V_{rstu}^{J 1})
\hat\delta_{rt}\hat\delta_{su}.
\label{(II.4b)}
\end{equation}
\end{mathletters}
There are two reasons.
 One is the Pauli principle, which operates when
$r=s$ or $t=u$, i.e. when $\delta (P)=1$ with $\delta(P)$ defined as
\begin{equation}
\delta (P)=1-(1-\delta_{rs})(1-\delta_{tu})
\label{(II.5)}
\end{equation}

A direct indication of the problem comes if we consider a single shell
with a constant interaction,$V_{rrrr}^\Gamma=1$. Then from
(\ref{(I.5)})
\[\sum Z_{rr\Gamma}^+\cdot Z_{rr\Gamma}=n_r(n_r-1)/2;\]
but from (\ref{(II.4a)}) we obtain
$\omega_{00}=(D_r-1)/2$ with $D_r=[r]$, and
inserting in (\ref{(II.3)}),
\[V_{\lambda=0}=( (D_r-1)/2D_r) n_r(n_r-1)/2\],
indicating that we miss the correct result by over a factor 2.  The
remaining half must be found among the $\omega^\lambda$ terms of
$V_{\lambda\not=0}$ , which in the case of $\delta (P)=1$ cannot be
linearly independent since they are twice as numerous as the
$V^\Gamma$ ones.  \medskip

The other problem is related to exchange terms that show for
$j_r=j_s=j_t=j_u$ and $\delta (P)=0$, i.e. when $\delta (e)=1$ with
\begin{equation}
\delta (e)\equiv (\hat\delta_{rs}-\delta_{rs})
(\hat\delta_{tu}-\delta_{tu})\,,
\label{(II.6)}
\end{equation}

Then, setting $V_{rsut}^\Gamma=1$ in (\ref{(II.4a)}) and
(\ref{(II.4b)}) and remembering that $V_{rsut}^\Gamma=-(-)^{J+T}
V_{rstu}^\Gamma$ (since $(-)^{2r}=1$ in $jt$ formalism)
\begin{mathletters}
\begin{equation}
\omega_{rust}^{00}=-[rs]^{-1/2}
\sum_{(J)} [J] (-)^J
(V_{rstu}^{J 0}-3V_{rstu}^{J 1})
\label{(II.7a)}
\end{equation}
\begin{equation}
\omega_{rust}^{01}=-[rs]^{-1/2}
\sum_{(J)} [J] (-)^J
(V_{rstu}^{J 0}+V_{rstu}^{J 1})
\label{(II.7b)}
\end{equation}
\end{mathletters}

indicating that much $\lambda=0$ monopole strength is hidden in the
$\omega_{rtsu}^{\lambda\tau}$ terms with $\lambda\not=0$.
\subsubsection{The form of ${\cal H}_{mT}$}
To achieve complete extraction we specify ${\cal H}_{mT}$ by {\it
  defect}, as suggested at the beginning of the section.  It is ${\cal
  H}_M$ that will be defined so as to contain no $V_{\lambda=0}$
terms. To illustrate how the prescription works we consider first the
$\delta(e)=0$ case which is simpler.

Since ${\cal H}_M$ has the form (\ref{(I.40)}), let us replace the
$V_{rstu}^{JT}$ matrix elements by
\begin{equation}
W_{rstu}^{JT}=V_{rstu}^{JT} -\overline  V_{rstu}^T
 \hat\delta_{rt}\hat\delta_{su}\quad{\rm if~}\delta(e)=0,
\label{(II.8)}
\end{equation}

with the ``centroids'' $\overline V_{rstu}^T$ defined so as to make
$\omega_{rtsu}^{0\tau}$ vanish:
\begin{equation}
  \hat\delta_{rt}\hat\delta_{su}\sum_{(J)}[J]
  W_{rstu}^{JT}=0\Longrightarrow\overline V_{rstu}^T=
  \sum_{(J)}[J]V_{rstu}^{JT} / \sum_{(J)}[J]
\label{(II.9)}
\end{equation}

Then ${\cal H}_{mT}$ is automatically defined in the $V$
representation as
\begin{equation}
{\cal H}_{mT}(\delta(e)=0)=K+
\sum_{r\leq s,\atop t\leq u,T}
\hat\delta_{rt}\hat\delta_{su}\overline  V_{rstu}^T
\sum_{(J)}Z_{rsJT}^+\cdot  Z_{tuJT}
\label{(II.10)}
\end{equation}

The remaining task is to write the operators in ${\cal H}_{mT}$ in
terms of $S_{rtsu}$ and $T_{rtsu}$, which we do easily by calling upon
eqs.(\ref{(I.47a)}) and (\ref{(I.47b)}), where
($\hat\delta_{rs}\hat\delta_{tu}=1$ of course)

\begin{eqnarray*}
\omega^{\lambda\tau}=\delta_{\lambda 0}\delta_{\tau 0}\Longrightarrow
V_{rstu}^{J 0}=V_{rstu}^{J 1} =[rs]^{-1/2}\\
\omega^{\lambda\tau}=\delta_{\lambda 0}\delta_{\tau 1}\Longrightarrow
V_{rstu}^{J 0}=3[rs]^{-1/2}\,,
V_{rstu}^{J 1} =-[rs]^{-1/2}
\end{eqnarray*}
which means
\begin{mathletters}
\begin{equation}
S_{rtsu}=\sum_{\Gamma}Z_{rs\Gamma}^+\cdot  Z_{tu\Gamma}
=\sum_JZ_{rsJ 0}^+\cdot  Z_{tuJ 0}+
\sum_JZ_{rsJ 1}^+\cdot  Z_{tuJ 1}
\label{(II.11a)}
\end{equation}
\begin{equation}
T_{rtsu}={3\over 4}\sum Z_{rsJ 0}^+\cdot  Z_{tuJ 0}-{1\over 4}
\sum Z_{rsJ 1}^+\cdot  Z_{tuJ 1}
\label{(II.11b)}
\end{equation}
\end{mathletters}
and inverting
\begin{mathletters}
\begin{equation}
\sum_JZ_{rsJ 0}^+\cdot  Z_{tuJ 0}={1\over 4}
(S_{rtsu}-4T_{rtsu})
\label{(II.12a)}
\end{equation}
\begin{equation}
\sum_JZ_{rsJ 1}^+\cdot  Z_{tuJ 1}={1\over 4}
(3S_{rtsu}+4T_{rtsu}).
\label{(II.12b)}
\end{equation}
\end{mathletters}
\medskip

The exchange contribution to ${\cal H}_{mT}$ is a bit trickier to
extract, because when $\delta(e)=1$, two extra operators have to be
considered.  They can be calculated from (\ref{(II.12a)}) and
(\ref{(II.12b)}) by exchanging $t$ and $u$,
\begin{mathletters}
\begin{equation}
\sum Z_{rsJ 0}^+\cdot  Z_{utJ 0}=
-\sum (-)^J Z_{rsJ 0}^+\cdot  Z_{tuJ 0}=
{1\over 4}(S_{rust}-4T_{rust})
\label{(II.13a)}
\end{equation}
\begin{equation}
\sum Z_{rsJ 1}^+\cdot  Z_{utJ 1}=
\sum (-)^J Z_{rsJ 1}^+\cdot  Z_{tuJ 1}=
{1\over 4}(3S_{rust}+4T_{rust})
\label{(II.13b)}
\end{equation}
\end{mathletters}
and by combining (\ref{(II.12a)}), (\ref{(II.12b)}),
(\ref{(II.13a)}) and (\ref{(II.13b)})
\begin{mathletters}
\begin{equation}
\sum_J Z_{rsJ 0}^+\cdot  Z_{tuJ 0}{(1\pm
(-)^J)\over 2}=
{1\over 8}( (S_{rtsu}-4T_{rtsu})\mp
(S_{rust}-4T_{rust}))
\label{(II.14a)}
\end{equation}
\begin{equation}
\sum_J Z_{rsJ 1}^+\cdot  Z_{tuJ 1}{(1\pm(-)^J)\over 2}=
{1\over 8}( (3S_{rtsu}+4T_{rtsu})\pm
(3S_{rust}+4T_{rust}))\quad
\label{(II.14b)}
\end{equation}
\end{mathletters}
Now it is possible to specify completely the matrix elements of ${\cal
  H}_M$, ($\delta(e)$ defined in (\ref{(II.6)})
\begin{equation}
W_{rstu}^{JT}= V_{rstu}^{JT}
-\hat\delta_{rt}\hat\delta_{su}\left\lbrack (1-\delta(e))
\overline  V_{rstu}^T+\delta(e) \overline  V_{rstu}^{\rho T}\right\rbrack\,,
\quad\rho=~{\rm sign~}(-)^J
\label{(II.15)}
\end{equation}

from which it follows that ${\cal H}_{mT}$ must be
\begin{eqnarray}
{\cal H}_{mT} =K+
\sum_{  r\leq s,\atop t\leq u,T,\rho=\pm}
\hat\delta_{rt}\hat\delta_{su}\left\lbrack (1-\delta(e))\overline  V_{rstu}^T
\Omega_{rstu}^T +\delta(e)
\overline  V_{rstu}^{\rho T}\Omega_{rstu}^{\rho T}
\right\rbrack\nonumber\\
\Omega_{rstu}^T =\sum_J Z_{rsJT}^+\cdot  Z_{tuJT}\qquad
\Omega_{rstu}^{\pm T} =\sum_J Z_{rsJT}^+\cdot  Z_{tuJT}\,
{(1\pm(-)^J)\over 2}
\label{(II.16)}
\end{eqnarray}
Through eqs.(\ref{(II.12a)}), (\ref{(II.12b)}), (\ref{(II.14a)}), and
(\ref{(II.14b)}) we can obtain the form of ${\cal H}_{mT}$ in terms of
the monopole operators by regrouping the coefficients affecting each
of them.  To simplify the presentation we adopt the following
convention

\[\left\{
\begin{array}{llll}
\alpha\equiv rstu&r\leq s,t\leq u,&
\hat\delta_{rt}\hat\delta_{su}=1,& {\rm BUT}\protect\\
S_\alpha=S_{rtsu},&S_{\bar\alpha}=S_{rust},&T_\alpha=T_{rtsu},&
T_{\bar\alpha}=T_{rust},
\end {array}
\right.\]

 then
\begin{mathletters}
\label{(II.17)}
\begin{eqnarray}
{\cal H}_{mT} =K+\sum_\alpha
(1-\delta(e))(a_\alpha S_\alpha+b_\alpha T_\alpha)\nonumber\\
+\delta(e) (a_\alpha^d S_\alpha+b_\alpha^d T_\alpha
+a_\alpha^e S_{\bar\alpha}+b_\alpha^e T_{\bar\alpha}),\,{\rm with}\\
a_\alpha ={1\over 4}(3\bar V^1_\alpha+\bar V_\alpha^0)\nonumber\\
b_\alpha ={1\over 4}( \bar V^1_\alpha-\bar V_\alpha^0)\\
a_\alpha^d ={1\over 8}(3\bar V_\alpha^{+1}+3\bar V_\alpha^{-1}
+\bar V_\alpha^{+0}+\bar V_\alpha^{-0})\nonumber\\
a_\alpha^e ={1\over 8}(3\bar V_\alpha^{+1}-3\bar V_\alpha^{-1}
-\bar V_\alpha^{+0}+\bar V_\alpha^{-0})\nonumber\\
b_\alpha^d ={1\over 2}({3}\bar V_\alpha^{+1}+{3}\bar V_\alpha^{-1}
-\bar V_\alpha^{+0}-\bar V_\alpha^{-0})\nonumber\\
b_\alpha^e ={1\over 2}({3}\bar V_\alpha^{+1}-\phantom{3}\bar
V_\alpha^{-1}
+\bar V_\alpha^{+0}-\bar V_\alpha^{-0})
\end{eqnarray}
\end{mathletters}

{}From (\ref{(II.9)}) and its generalization to $\delta(e)=1$ we find
for the centroids
\begin{eqnarray}
\overline  V_{rstu}^T=\sum_{(J)} V_{rstu}^{JT}[J]
/ \sum_{(J)}[J] \qquad
\sum_{(J)}[J] ={1\over 4}\;{D_r(D_s+2
\delta(P)(-)^T)\over 1+\delta(P)}\nonumber\\
\overline  V_{rstu}^{\pm T}=\sum_J V_{rstu}^{JT}[J]
(1\pm(-)^J)/ \sum_J[J] (1\pm(-)^J)\nonumber\\
\sum_J[J](1\pm (-)^J) ={1\over 4}\,D_r(D_r\mp 2)\nonumber\\
D_r=[r],\quad \hat\delta_{rt}\hat\delta_{su}=1\,{\rm always; }\quad
\delta(e)=1,\;{\rm for~}
V_{rstu}^{\pm T}\,.\label{(II.18)}
\end{eqnarray}

\subsubsection{Form of ${\cal H}_{mnp}$ and ${\cal H}_{m0}$  }

In $j$ formalism ${\cal H}_{mnp}$ is ${\cal H}_{mT}$ under another
guise: neutron and proton shells are differentiated and the operators
$T_{rs}$ and $S_{rs}$ are written in terms of 4 scalars
$S_{r_xs_y}\;\;; x,y=n \;{\rm or} \;p$.

We may also be interested in remaining in $jt$ formalism and extract
only the purely isoscalar contribution to ${\cal H}_{mT}$, which we
call${\cal H}_{m0}$.  The power of French's product notation becomes
particularly evident here, because the form of both terms is
identical. It is left as an exercise to find
\begin{equation}
{\cal H}_{mnp}\;or\;{\cal H}_{m0} =K+\sum_\alpha \left\lbrace
\bar V_\alpha S_\alpha(1-\delta(e))+{1\over 2}
((\bar V_\alpha^++\bar V_\alpha^-)S_\alpha+
(\bar V_\alpha^--\bar V_\alpha^+)S_{\bar\alpha})\delta(e)\right\rbrace
\label{(II.19)}
 \end{equation}

with
\begin{eqnarray}
\overline  V_{rstu}=\sum_{(\Gamma)} V_{rstu}^\Gamma[\Gamma]/
\sum_{(\Gamma)}[\Gamma] \nonumber\\
\sum_{(\Gamma)}[\Gamma] =D_r(D_s-\delta(P))/
( 1+\delta(P))\nonumber\\
 \overline  V_{rstu}^\pm=\sum_{\Gamma} V_{rstu} [\Gamma]
(1\pm(-)^{\Gamma+2r})\quad\nonumber\\
\sum_{\Gamma}[\Gamma](1\pm(-)^{\Gamma+2r})=D_r(D_r\mp(-)^{2r})
\label{(II.20)}
\end{eqnarray}
Of course we must remember that for ${\cal H}_{mnp}$
$D_r=2j_r+1,\;(-)^{2r}=-1,\;\Gamma\equiv J,$ etc., while for ${\cal
  H}_{m0}$

$D_r=2(2j_r+1),\;(-)^{2r}=+1,\;\Gamma\equiv JT$; etc.

It should be noted that ${\cal H}_{m0}$ is not obtained by simply
discarding the $b$ coefficients in eqs.(\ref{(II.17)}) , because we
can extract some $\gamma=$00 contribution from the $T_\alpha$
operators. The point will become quite clear when considering the
diagonal contributions.

\subsection{ Diagonal forms of $  {\cal H}_m$}

We are going to specialize to the diagonal terms of ${\cal H}_{mT}$
and ${\cal H}_{mnp}$, which involve only $V_{rsrs}^\Gamma$ matrix
elements whose centroids will be called simply $V_{rs}$ and $V_{rs}^T$
(the overline in $\overline V_{rstu}^T$ was meant to avoid confusion
with possible matrix elements $V_{rstu}^1$ or $V_{rstu}^0$, it can be
safely dropped now).  Then (\ref{(II.17)}) and (\ref{(II.19)}) become
\begin{mathletters}
\begin{equation}
{\cal H}_{mT}^d =K+\sum_{r\leq s}( a_{rs}\,n_r(n_s-\delta_{rs})
             /(1+\delta_{rs})+b_{rs}(T_r\cdot  T_s-{3\over 4}n
 \delta_{rs})/(1+\delta_{rs}))
\label{(II.22a)}
\end{equation}
\begin{equation}
{\cal H}_{mnp}^d\;or\;{\cal H}_{m0}^d =
K+\sum_{r\leq s}   V_{rs}\,n_r(n_s-\delta_{rs}) /(1+\delta_{rs})
\label{(II.22b)}
\end{equation}
\end{mathletters}

We rewrite the relevant centroids incorporating explicitly the Pauli
restrictions
\begin{mathletters}
\begin{equation}
V_{rs}={\sum_\Gamma V_{rsrs}^\Gamma[\Gamma]
( 1-(-)^\Gamma\delta_{rs})\over D_r(D_s-\delta_{rs})}\quad
V_{rs}^T={4\sum_J V_{rsrs}^{JT}[J]
( 1-(-)^{J+T}\delta_{rs})\over D_r(D_s+2\delta_{rs}(-)^T)}
\label{(II.23)}
\end{equation}
\begin{equation}
a_{rs}={1\over 4}(3V^1_{rs}+V^0_{rs})=V_{rs}+{3\over 4}\,
{\delta_{rs}\over D_r-1}\,b_{rs}\quad
b_{rs}=V^1_{rs}-V^0_{rs}\label{(II.24)}
\end{equation}
\end{mathletters}
The relationship between $a_{rs}$ and $V_{rs}$ (which should be
checked) makes it possible to combine eqs.(\ref{(II.22a)}) and
(\ref{(II.22b)}) in a single form
\begin{equation}
{\cal H}_m^d =K+\sum V_{rs}\,n_r(n_s-\delta_{rs})
 /(1+\delta_{rs})+b_{rs}\left(T_r\cdot  T_s-{3
n_r\bar n_r\over 4(D_r-1)}\delta_{rs}\right)/(1+\delta_{rs})
\label{(II.25)}
\end{equation}

in which {\it now} the $b_{rs}$ term can be dropped to obtain ${\cal
  H}_{m 0}^d$ {\it or} ${\cal H}_{mnp}^d$.  Note that the diagonal
terms depend on the representation:

${\cal H}_{mnp}^d\not ={\cal H}_{mT}^d$ in general.  \medskip It is
useful to have at hand the relationship between matrix elements and
centroids in the $jt$ and the $j$ schemes. A single shell $r$ in the
former becomes a pair $r_n$ and $r_p$ in the latter

\begin{eqnarray}
V_{r_ns_pt_nu_p}={(1+\delta_{rs})^{1/2}(1+\delta_{tu})^{1/2}\over 2}\;
( V_{rstu}^{J1}+V_{rstu}^{J0})\nonumber\\
V_{r_ns_pu_pt_n}={(1+\delta_{rs})^{1/2}(1+\delta_{tu})^{1/2}\over 2}\;
( V_{rsut}^{J1}-V_{rsut}^{J0})\nonumber\\
V_{r_ns_nt_nu_n}=V_{r_ps_pt_pu_p}=V_{rstu}^{J1}
\label{(II.26)}
\end{eqnarray}
and for the centroids ($x,y=n$ or $p$, $x\not=y$)
\begin{equation}
V_{r_xs_x}=V_{rs}^1\qquad
V_{r_xs_y}={1\over 2}\left\lbrack V_{rs}^1\left(1-{2\delta_{rs}\over
      D_r}
\right)+ V_{rs}^0\left(1+{2\delta_{rs}\over D_r}
\right)\right\rbrack
\label{(II.27}
 \end{equation}

\subsubsection{Trace equivalence and propagation}

We have already mentioned the important - and known \cite{FR69}-
property of ${\cal H}_m^d$ that its expectation value for a given
state reproduces the trace of ${\cal H}$ over configurations
$Q=(n_1\,n_2\,\ldots n_r\ldots)$ or $Q=(n_1T_1\,n_2T_2\,\ldots
n_rT_r)$, to which the state belongs. The notation $Q$ stands for
fixed $n_r$ or $n_rT_r$ in each subshell.  Averages at fixed seniority
\cite{DST63} and for other symmetries are also of much interest
\cite{PA78}.  Here we propose a very general proof that relies on
Schur's lemma and on a trick used by French \cite{FR69}. Let now $Q$
stand for any symmetry associated with states $| YQ>$ of $n$ particles
($n$ is included in $Q$).  The trick is to construct $| YQ>$ by acting
on $|\bar0>$ instead of $| 0>$. Let us introduce ${\cal D}_{\bar Q}$
of rank $\bar n=D-n$,
\begin{equation}
{\cal D}_{\bar Q}=\sum_YZ_{Y\bar Q}^+Z_{Y\bar Q},
\quad| YQ>=Z_{Y\bar Q}|\bar 0>,
\quad| Y\bar Q>=Z_{Y\bar Q}^+| 0>\,;
\label{(II.28a)}
\end{equation}

The quantum numbers $\bar Q$ are associated to the ph transforms of
the operators e.g.  $\bar Q=(\bar n_1T_1\,\bar n_2T_2\,\ldots)$.
${\cal D}_{\bar Q}$ is a projector
\begin{equation}
{\cal D}_{\bar Q}| Y\bar Q'>=\delta_{QQ'}| Y\bar Q>
\quad{\rm if}\quad \bar n'\leq \bar n
\label{(II.28b)}
\end{equation}

{}From Schur's lemma it follows that it can be written in terms of the
Casimir operators associated with the $Q$ quantum numbers. It is seen
that ${\cal D}_{\bar Q}$ is a generalization of ${\cal D}_k$ in
(\ref{(I.5)}) and for $Q=(n_1\,n_2\,\ldots n_r\ldots)$, it is obvious
that
\begin{equation}
{\cal D}_Q={\cal D}_{n_1\,n_2\,\ldots n_r\ldots}=\prod_{r}
{\hat n_r\choose n_r}\,;\qquad
{\cal D}_{\bar Q}={\cal D}_{\bar n_1\,\bar n_2\,\ldots \bar n_r}
\label{(II.29)}
\end{equation}

where we are forced to distinguish the operator $\hat n_r$ and its
eigenvalue $n_r$. Note that ${\cal D}_{\bar Q}$ is {\it not} the ph
transform of ${\cal D}_Q$.

We call $d_Q=\sum_Y< YQ| YQ>=<\bar 0| {\cal D}_{\bar Q}|\bar 0>$; the
dimension of the space over which we take the trace $<{\cal H}>_Q$;
and $W_{xxq}Z_{xq}^+Z_{xq}$ the diagonal contributions to ${\cal H}$
(in our case $x=J,\;q=rsT$). Now we can calculate the trace (a step by
step explanation follows the equations):
\begin{mathletters}
\label{(II.30)}
\begin{equation}
<{\cal H}>_Q =
d_Q^{-1}\sum_Y
< YQ|\sum_{xq}W_{xxq}Z_{xq}^+Z_{xq}| YQ>=
\end{equation}
\begin{equation}
=d_Q^{-1}\sum_{xqY}W_{xxq}<\bar 0|
Z_{Y\bar Q}^+Z_{xq}^+Z_{xq}Z_{Y\bar Q}| \bar 0>=
\end{equation}
\begin{equation}
=d_Q^{-1}\sum_{xq}W_{xxq}<\bar 0|
Z_{xq}^+{\cal D}_{\bar Q}Z_{xq}| \bar 0>=
\end{equation}
\begin{equation}
=d_Q^{-1}\sum_q\bigg\lbrack< \bar 0|
Z_{x'q}^+{\cal D}_{\bar Q}Z_{x'q}| \bar 0>
\sum_xW_{xxq}\bigg\rbrack=
\end{equation}
\begin{equation}
=d_Q^{-1}\sum_q\bigg\lbrack d_q^{-1}\sum_{x'}< QY|
Z_{x'q}^+Z_{x'q}| QY> \sum_xW_{xxq}\bigg\rbrack=
\end{equation}
\begin{equation}
=< \sum_q{\cal D}_qd_q^{-1}
\sum_xW_{xxq}>_Q.
\end{equation}
\end{mathletters}
Line by line we have

a. definition of trace

b. insert the definition of $| YQ>=Z_{Y\bar Q}|\bar 0>$

c. French's trick of using $Z_{Y\bar Q}| \bar 0>$ instead of
$Z_{YQ}^+| 0>$ for $| YQ>$ allows to permute operators.
 Then use eq.(\ref{(II.28a)})

d. the crucial step: since the expectation value of ${\cal D}_{\bar Q}$
is independent of $x$ we take it out of the sum
for some arbitrary $x'$.

e. by the same token we sum over $x'$ and divide by the number of terms,
then commute again the operators in $q$ and $Q$.

f. rearrange e to make clear what the {\it trace equivalent}
operator is \cite{FR69}.

If we specialize this result to our problem, eqs.(\ref{(II.12a)}) and
(\ref{(II.12b)}) for $r=t,\;s=u$ provide ${\cal D}_q$ and
eq.(\ref{(II.9)}) provides $d_q^{-1}\sum W_{xxq}$. Therefore, ${\cal
  H}_{mT}^d$ is indeed the trace equivalent of ${\cal H}$.  \medskip
Reference to Schur's lemma may be avoided for our particular problems.
What has to be shown is that we have exactly the number of operators
required to calculate ${\cal D}_Q$. For ${\cal H}_{mnp}^d$ the problem
is solved in (\ref{(II.29)}).  For ${\cal H}_{mT}^d$ the form of
${\cal D}_Q$ is not so simple, but we can easily check that the number
of operators available meets exactly our needs.  Let us do it for
1~shell (or for total $n$ and $T$), so that ${\cal D}_q={\cal
  D}_{nT}$.

For $n=2$ we have from eqs.(\ref{(II.12a)}) and (\ref{(II.12b)}) ,
$(T^2\equiv T(T+1))$,
\begin{equation}
{\cal D}_{20}={1\over 8}\,(n(n+2)-4T^2)\qquad
{\cal D}_{21}={1\over 8}\,(3n(n-2)+4T^2)
\label{(II.31)}
\end{equation}

For $n=3$ we need ${\cal D}_{3 {1\over2}}$ and ${\cal D}_{3
  {3\over2}}$ which we construct out of $n^{(3)}$ and
$(n-2)\,(T^2-3/4n)$, the two available rank~3 operators.  For $n=4$ we
need ${\cal D}_{40},\;{\cal D}_{41},\;{\cal D}_{42}$ and we have
$n^4,\; n^2\,T^2\,T^4$ from which we can extract rank~4 operators and
work out
\begin{eqnarray}
{\cal D}_{40}={1\over 12}\left(T^2-{n(n+2)\over 4}\right)
\left(T^2-{1\over 4}n(n+2)\right)\nonumber\\
{\cal D}_{41}={1\over 8}\left(T^2-{n(n+2)\over 4}\right)
\left(T^2-{3\over 4}(n-2)(n-4)\right)\nonumber\\
{\cal D}_{42}={n^{(4)}\over 24}-{\cal D}_{40}-{\cal D}_{41}
\label{(II.32)}
\end{eqnarray}
Obviously the construction can be extended to any ${\cal D}_{n T}$ .

\medskip

If we want to deal with averages at fixed intermediate $T$
values e.g.

$Q\equiv (n_1T_1\,n_2T_2\,n_3T_3\;T_{12}\;T)$, even the counting may
become hard: Schur's lemma is not trivial. Still we can do without it,
by direct averaging through standard multipole techniques, as in
\begin{eqnarray}
\sum_{Y_1Y_2J_1J_2J}< n_1\Gamma_1Y_1n_2\Gamma_2Y_2\,\Gamma\|
\left\lbrack(A_1B_A)^\gamma(A_2B_2)^\gamma\right\rbrack^0\|
n_1\Gamma_1Y_1n_2\Gamma_2Y_2\,\Gamma>=\nonumber\\
=\sum_{Y_1Y_2J_1J_2J}
(-)^{\Gamma_1+\Gamma_2-\Gamma-\gamma}\left\lbrace
\matrix{\Gamma_1\,\Gamma_2\,\Gamma\cr
\Gamma_2\,\Gamma_1\,\Gamma\cr}\right\rbrace
< \Gamma_1Y_1\|  S_1^\gamma\|  \Gamma_1Y_1>
< \Gamma_2Y_2\|  S_2^\gamma\|  \Gamma_2Y_2>
\label{(II.33)}
\end{eqnarray}
where, using the sum rule (A19) for $J$, only $\gamma=01$ and 00
survive, which are absent from ${\cal H}_M$, establishing ${\cal H}_m$
as trace equivalent for these operators. The only averages that cannot
be done in this way are the basic ones involving $n$ and $T$ only, but
these we have already.

The idea to associate a trace to an equivalent operator and not only
to a number, leads to French's idea of {\it propagation}. It stresses
the fact that by combining the general properties and particular
symmetries of a given trace equivalent operator we may {\it determine}
it at some convenient and limited set of $Q$ values and then {\it
  propagate} i.e. apply it to any $Q$. The task is hard in general but
for purely isocalar averages it is often possible to to guess outright
the solution, as for the ``width'' of ${\cal H}_M$:
\begin{equation}
\sigma^2(n)=<{\cal H}_M^2>_n={n^{(2)}\bar n^{(2)}\over
2(D-2)^{(2)}}\sigma^2(2)\,,\quad  \sigma^2(2)=
{2\over D^{(2)}}\sum W_{ijk\ell}^2
\end{equation}
since the trace equivalent operator must be of rank 4, reproduce
$\sigma^2(2)$ at $n=2$, and vanish at $n=0,1$, $\bar n=0,1$ because
the energies of the corresponding states are entirely given by $ {\cal
  H}_m$ .

\subsection{ The Hartree-Fock property of $  {\cal H}_m$. Strict
rank operators}

Equation (\ref{(I.38)}) contains all the possible contractions that
may appear in any possible change of vacuum.  By construction they all
vanish for ${\cal H}_M$ since their coefficients are proportional to
centroids that have all migrated to ${\cal H}_m$. As a consequence
${\cal H}_M$ is of rank~2 both for particles and holes (we call this
property, strict rank-2) and its expectation value vanishes for {\it
  any} closed shell and {\it any} single particle or single hole
state, a set we denote by ``$CS\pm1$".  As a consequence,
\begin{equation}
< CS\pm 1|{\cal H}| CS\pm 1>=
< CS\pm 1|{\cal H}_m| CS\pm 1>
\label{(II.34)}
\end{equation}

which remains valid under unitary transformations $a_r^+=\sum
U_{rx}b_x^+$ because centroids become linear combinations of
centroids. Therefore, if they all vanish in one representation, so
they do in any other representation and ${\cal H}_m$ remains a strict
rank-2 operator.  Of course at any particular basis we can replace
${\cal H}_m$ by ${\cal H}_m^0$ in (\ref{(II.34)}).

{}From these remarks it follows that

\smallskip

{\em Hartree-Fock variation of ${\cal H}$ is Hartree-Fock
variation of ${\cal H}_m$.}

\smallskip

Eq. (\ref{(II.34)}) can be obtained more directly by arguing that any
$CS\pm 1$ state is the only member of the configuration to which it
belongs.  If we have invoked eq.(\ref{(I.38)}) instead, it is to
introduce the concept of strict rank-2 operators. It can be certainly
generalized and it is quite useful for monopole operators, although it
must take a weaker form since for ${\cal H}_M$ the property guarantees
vanishing for all $CS\pm 1$ states, which cannot be the case for any
diagonal contribution ${\cal H}_m$.  For instance: in eq.
(\ref{(II.25)}) the isoscalar part has been cleanly separated because
$T^2-3/4 n$ of rank~2 in eq.(\ref{(II.22a)}) has been changed into
$T^2-{3n\bar n\over 4(D-1)}$ which is of strict rank~2 for doubly
closed shells.

For number operators it is easy to check that
\begin{equation}
\Gamma_r=\left({n_r\over D_r}-{n_1\over D_1}\right) D_r=
-\left({\bar n_r\over D_r}-{\bar n_1\over D_1}\right) D_r=
-\bar\Gamma_r
\label{(II.35)}
\end{equation}

is of strict rank~1 for closed shells containing 1 and $r$, while
\begin{equation}
\Gamma_{rs}=\left({2n_rn_s\over D_rD_s}-{n_r^{(2)}\over D_r^{(2)}}
-{n_s^{(2)}\over D_s^{(2)}}\right) {D_rD_s\over 2}=
\bar\Gamma_{rs}\equiv\Gamma_{rs}(\bar n_r,\bar n_s)
\label{(II.36)}
\end{equation}

is of strict rank~2 for closed shells containing $r$ and $s$.

It is seen that the set of operators $n_1,\,n_2,\,\ldots n_\nu$ can be
replaced by the total number $n$ and $\nu-1$ operators $\Gamma_r$ (in
which the choice of $n_1$ is arbitrary). Similarly we have $n^{(2)},\,
n\Gamma_r$ and $\Gamma_{rs}$ with $r\not=1$ and $r\not=s$, replacing
the $n_r\,n_s$ set.

The form of ${\cal H}_{m 0}^d$ obtained with this construction is
given in \cite{ACZ91} where it made possible an efficient monopole
phenomenology (beware of two erratae in this reference: in the
definition of $\epsilon_r$ read $N_r$ not $N_{ir}$, and in that of
$W_r$ read $N_s$ not $N_{r}$).  For the form of ${\cal H}_{mnp}^d$
refer to \cite{ZU94}, where several applications are given. In
particular a reinterpretation of BCS as a number conserving theory
which hinges on what is the fundamental point of the construction:
That it singles out $n$ - always a conserved quantity, and the only
one of strict rank 0 - from all the other number operators.

To understand the interest of this result in giving a full
characterization of ${\cal H}_{m 0}$ we note that the energy of a
closed shell to which the $n\Gamma_r$ and $\Gamma_{rs}$ operators do
not contribute, is given by $Wn(n-1)/2$, with
\[W=\sum_{}V_{rs}D_{rs}/\sum{}D_{rs} ~~~
D_{rs}=D_r(D_s-\delta_{rs})/(1+ \delta_{rs}).\]
It means that a single parameter $W$ is in principle sufficient to
describe the energies of all closed shells. The catch is that all the
strict rank properties apply to the particular closed shell being
considered and $W$ is necessarily space-dependent, since defined only
for a given set of occuppied orbits. In a way we have gained a lot by
isolating the strongest contributor but it is of no use to have a
parameter that is space dependent if we do not know the dependence. In
\cite{DZ}, we shall see how the problem is naturally solved, once we
consult the realistic interactions.  \medskip

The separation property of ${\cal H}_{m }$ had been anticipated
\cite{ZU75}, used\cite{ACZ91}, and very sketchily proved for ${\cal
  H}_{mnp}$\cite{ZU94}.  The full derivation was long overdue and it
might have been given in a briefer form but we hope that the longer
one we have chosen will be more useful.

\appendix
\section{}

{\bf  Reduced matrix elements for
$  \ell,\,\sigma,\,rY_1$ and $  q$ operators}

Conventions: radial wavefunctions are positive near the origin.

$\widehat j=\widehat\ell+\widehat s$ (not $\widehat s+\widehat\ell$).

Condon and Shortley phases for $Y_{\ell m}$.

\bigskip

A) $\widehat\sigma=2\widehat s$ and $\widehat\ell$ operators.

\begin{eqnarray}
j'=\ell+1/2\qquad j=\ell-1/2\qquad f(j)=( j(j+1)(2j+1))^{1/2}
\nonumber\\
< j'\| \widehat\sigma\|  j'>=f(j')/ j'\qquad
< j\| \widehat\sigma\|  j>=-f(j)/(j+1)
\label{(A.1)}\\
< j'\| \widehat\ell\|  j'>=\ell f(j')/ j'\qquad
< j\| \widehat\ell\|  j>=(\ell+1)f(j)/(j+1)
\label{(A.2)}\\
< j\| \widehat s\|  j'>=< j'\| \widehat\ell\|  j>
=\left\lbrack{2\ell(\ell+1)\over 2\ell+1}\right\rbrack^{1/2}
=-< j'\| \widehat s\|  j>=-< j\| \widehat\ell\|  j'>
\label{(A.3)}
\end{eqnarray}

\bigskip

B) $r^\lambda Y_\lambda$ operators.
$\quad Y_\lambda=\left({2\ell+1\over 4\pi}\right)^{1/2}\,C_\lambda$
\begin{equation}< p\ell j\|  r^\lambda C_\lambda\|  p'\ell' j'>=
< p\ell| r^\lambda | p'\ell'> (-)^{j'-\lambda-1/2}
[jj']^{1/2}\pmatrix{j&j'&\lambda\cr \cr 1/2&-1/2&0\cr}
\label{(A.4)}   \end{equation}

\bigskip

B1) $r^1 Y_1$ operators.
\begin{eqnarray}
< p\ell| r| p+1\;\ell+1>=\sqrt{{1\over 2}(p+\ell+3)}
\label{(A.5)}\\
< p\ell| r| p+1\;\ell-1>=-\sqrt{{1\over 2}(p-\ell+2)}
\label{(A.6)}\\
< p\ell j\|  r C_1\|  p+1\;\ell+1j+1>=-\sqrt{{p+\ell+3\over 2}}
\left\lbrack{(2j+1)(2j+3)\over 2(2j+2)}\right\rbrack^{1/2}
\label{(A.7)}\\
< p\ell j\|  r C_1\|  p+1\;\ell+1\; j>=-\sqrt{{p+\ell+3\over 2}}
\left\lbrack{2j+1\over 2j(2j+2)}\right\rbrack^{1/2}
\label{(A.8)}\\
< p\ell j\|  r C_1\|  p+1\;\ell-1\; j>= \sqrt{{p-\ell+2\over 2}}
\left\lbrack{2j+1\over 2j(2j+2)}\right\rbrack^{1/2}
\label{(A.9)}\\
< p\ell j\|  r C_1\|  p+1\;\ell-1\;
j-1>=-\sqrt{{p-\ell+2\over 2}}
\left\lbrack{(2j-1)(2j+1)\over 4j}\right\rbrack^{1/2}\quad
\label{(A.10)}
\end{eqnarray}

\bigskip

B2) $r^2 Y_2$ operators ($q_{0\hbar\omega}$ only)
\begin{eqnarray}
< p\ell| r^2| p\ell>=p+3/2\qquad
< p\ell| r^2| p\ell+2>=-( (p-\ell)(p+\ell+3))^{1/2}
\label{(A.11)}\\
< j\ell\|  r^2 C_2\|  j\ell>=-{p+3/2\over 2}
\left\lbrack{(2j+1)(2j-1)(2j+3)\over 2j(2j+2)}\right\rbrack^{1/2}
\label{(A.12)}\\
< j\ell\|  r^2 C_2\|  j+1\;\ell>=(p+3/2)
\left\lbrack{3\over 2}\,{(2j+1)(2j+3)
\over 2j(2j+2)(2j+4)}\right\rbrack^{1/2}
\label{(A.13)}\\
< j\ell\|  r^2 C_2\|  j+1\;\ell+2>=
-\left\lbrack{3\over 2}\,{(p-\ell)(p+\ell+3)(2j+1)(2j+3)
\over 2j(2j+2)(2j+4)}\right\rbrack^{1/2}
\label{(A.14)}\\
< j\ell\|  r^2 C_2\|  j+2\;\ell+2>=
-\left\lbrack{3\over 8}\,{(p-\ell)(p+\ell+3)(2j+5)(2j+3)(2j+1)
\over (2j+4)(2j+2)}\right\rbrack^{1/2}
\label{(A.15)}
\end{eqnarray}

{\bf  Useful formulae involving
$  3j$ and $  6j$ symbols}
\begin{equation}
\left.\matrix{< jm\,j'm'\,| JM>=(-)^{j-j'+M}[J]^{1/2}\pmatrix{j&j'&J\cr
m&m'&-M\cr}=\hfill\cr\cr
=(-)^{j+j'-J}< j'm'\,jm| JM>=
(-)^{j+j'-J}< j-mj'-m'| J-M>\cr}\right\rbrace\label{(A.16)}
 \end{equation}

\begin{equation}
\pmatrix{j&j'&J\cr  m&m'&-M\cr}=(-)^{j+j'+J}
\pmatrix{j'&j&J\cr  m'&m&-M\cr}=(-)^{j+j'+J}
\pmatrix{j&j'&J\cr -m&-m'&M\cr}\label{(A.17)}
   \end{equation}

The $3j$ symbol is invariant under cyclical permutations.
\begin{equation}< jm\,j-m| 00>=[j]^{-1/2} (-)^{j-m}
\label{(A.18)}   \end{equation}

\begin{equation}
\sum_J(-)^J[J] \left\lbrace\matrix{j&j'&J\cr j'&j&k\cr}
\right\rbrace=(-)^{j+j'}[jj']^{1/2}\delta_{k0}
\label{(A.19)}
 \end{equation}

\begin{equation}
\sum_J[J] \left\lbrace\matrix{j&j'&J\cr j&j'&k\cr}\right\rbrace=1
\label{(A.20)}
 \end{equation}

\begin{equation}
\left\lbrace\matrix{j&j&0\cr j'&j'&k\cr}\right\rbrace=
(-)^{j+j'+k}[jj']^{-1/2}
\label{(A.21)}
\end{equation}

The $6j$ symbol is invariant under permutations of columns and under
interchange of upper and lower indices in two columns.  \medskip

{\bf  NOTATIONS}

We collect here a list of notations used throughout the text,
indicating, when useful, the equation or section in which they are
introduced or explained. Occasionally some comments are made.
\medskip

WARNING: The same notation may mean different things, but once
this is kept in mind it should be very easy to give the correct
interpretation in context .

{\bf Three general points}

1) $F^{(n)}=F(F-1)\cdot s (F-n+1)$, integer $n$. Boole's factorial
power notation.
\medskip

2) Hardy's Order notation $f(x)=O(x)$ means $f(x)/x\,<$~constant,
for $x$ tending to some limit ($x\to\infty$ or $x\to0$ usually).
\medskip

3) $T_\kappa^k=T_{k\kappa}$ (never $T^{k\kappa}$) is a tensor operator
of rank $k$ and projection $\kappa$. Applies to coupled operators as
in $(T^kS^{k'})_{K_z}^K$, in which case the rank $K$ is {\it always}
a superindex. Whenever possible, projections are omitted. Zero
coupling: $(T^kS^k)^0$, when working in product space ($JT,\;LS$ or
$LST$), means full scalar. E.g. in $JT$ space
$0=00$, i.e. scalar, isoscalar.
\medskip

{\bf Symbols}

\begin{itemize}

\medskip

\item{$\Longrightarrow$} implies; $\to$ transforms, tends to, becomes.
\medskip

\item{$[j]$} $=2j+1$, but see (\ref{(I.11)}) for product spaces. Also
$[{\bf i}]=i_1,\,i_2,\ldots i_n$, but only in the very first lines of
section~I.
\medskip

\item{$<~>_Q$} $=$ trace (\ref{(II.30)}).
\medskip

\item{$\cdot $} scalar product $\quad T\mathop{\cdot }\limits^k(S^k)^*=
\sum_\kappa T_\kappa^k(S_\kappa^k)^*$
\medskip

\item{$^*$} Hermitean conjugation for operators, complex conjugation
for numbers.
\medskip

\item{$+$} Creation symbol as in $a^+$, $Z^+$, $X^+$ which are creation
operators.
\medskip

\item{$-$} ph transform (\ref{(I.6)})\hfill\break
Adjoint (\ref{(I.18)})\hfill\break
Centroid (\ref{(II.18)})\hfill\break

\medskip

\item{$\widehat{~}$} $n,\,J,\,T$ are used as operators and eigenvalues. When
necessary to distinguish them we use $\widehat n,\,\widehat J$ and
$\widehat T$ for
the operators.
\medskip

\item{$\sum_{(~)}$} restricted sum (\ref{(II.14a)}) and (\ref{(II.14b)}).
\medskip
\end{itemize}

{\bf Acronyms}

CM: center of mass.

CS, CS$\pm$1: Closed shell, CS plus 1 particle and 1 hole states.

HF: Hartree Fock.

HO: Harmonic oscillator.

NN: Nucleon-Nucleon.

$np$ or $\nu\pi$: neutron, proton.

ph: particle hole.

rhs and lhs: right hand side, left hand side.

SM: Shell Model.
\medskip

{\bf Operators, variables}

$a_r^+,\,a_r$ (1.1), $A_r,\,B_R$ (\ref{(I.23)}), creation, annihilation
operators

$A$: total number of particles.
${\cal A}_{rs}^\gamma$: antisymmetric multipole (\ref{(I.30)})

$d$: dimension of space.

$D=\sum_r D_r=\sum [r]$. Degeneracies
of space and orbit $r$.

$G$: pairing strength.

${\cal H}$: Hamiltonian in full space.

$H$: Hamiltonian in subspace.

$i,\;j,\;k,\;\ell$: individual orbits .

$J,\; j_r$: angular momentum (total or of {\it orbit} $r$).

$K$: kinetic energy.

$L,\;\ell_r$: orbital angular momentum (total, or of orbit $r$).

$M$: nucleon mass.

$n,\;n_r$: number of particles (total or for shell $r$),
$\bar n,\;\bar n_r$: number of holes.

$N$: number of neutrons

$p,\;p_r$: principal quantum number $p_r=2\nu_r+\ell_r$.

$p$: momentum (see under $r$).

$P_{rs}$ convention (\ref{(I.39)}).

$r,s,t,u$: subshells. $r\equiv (j_r p_r)$.

$S_{rs}^\gamma=S_{\gamma_z}^\gamma(rs)=(A_rB_s)_{\gamma_z}^\gamma$
(\ref{(I.24)}) $\quad {\cal S}_{rs}^\gamma$ (\ref{(I.30)}).

$S_{rs},\;T_{rs}$  (\ref{(II.1)});$\quad S_{rstu},\;T_{rstu}$
 (\ref{(II.2a)}),(\ref{(II.2b)}).

$T$: isospin.

$V$: potential energy. $\quad V$ representation section I.B

$V_{rstu}^\Gamma,\; V_{ijk\ell}$, two body matrix elements (I.16).

$\overline  V_{rstu}$ (\ref{(II.18)})(\ref{(II.20)})
 $V_{rs}$ (\ref{(II.23)}), centroids.

$W$ representation (I.40).

$W_{rstu}^\Gamma$ (II.8).

$X_{\Gamma\Gamma_z}^+(rs)=(A_rA_s)^\Gamma_{\Gamma_z},\quad
X_{\Gamma\Gamma_z}(rs)=(B_rB_s)^\Gamma_{\Gamma_z}$ (\ref{(I.24)}).

$Z_{\Gamma\Gamma_z}(rs)$ or
$Z_{rs\Gamma}^+$ (\ref{(I.14)}).

$Z$: number of protons.

$\alpha\equiv rstu$ (i.e.
$W_{rstu}^\Gamma=W_\alpha^\Gamma$)

$\gamma\equiv \lambda\tau$, $\Gamma=JT$ product tensor indices in
the $V$ and $omega$ representations.

$\delta_{rs}$ Kronecker $\delta$.

$\hat\delta_{rs}=\delta j_rj_s\,\delta \pi_r\pi_s\,,\;\pi_r=(-)^{p_r}$.

$\zeta_{rs}=(1+\delta_{rs})^{-1/2}$

$\lambda,\tau$: angular momentum and isospin in
$\omega$ representation.

$\nu_r=$ nodal quantum number: $p_r=2\nu_r+\ell_r$.

$\nu,\;\pi$: neutron, proton.

$\tau_r$: isospin projection for orbit $r$.

$\omega_{rstu}^\gamma$: (\ref{(I.41)},\ref{(I.42a)},\ref{(I.42b)}).

$\Omega_r=j_r+1/2\quad \Omega=\sum_r \Omega_r$

$\omega=$ HO frequency.

\vfill\eject

\end{document}